\documentclass[reprint,amsmath,amssymb,floatfix,aps]{revtex4-2}

\usepackage{graphicx}
\usepackage{dcolumn}
\usepackage{bm}

\usepackage{cmap}					
\usepackage{mathtext} 				
\usepackage[T2A]{fontenc}			
\usepackage[utf8]{inputenc}			
\usepackage[english,russian]{babel}	

\graphicspath{{images/}{images/}}  

\renewcommand{\epsilon}{\ensuremath{\varepsilon}}
\renewcommand{\phi}{\ensuremath{\varphi}}
\renewcommand{\kappa}{\ensuremath{\varkappa}}

\renewcommand{\vec}[1]{\ensuremath\bm{#1}}
\renewcommand{\eqref}[1]{Eq.~(\ref{#1})}
\newcommand{\fref}[1]{Fig.~\ref{#1}}

\usepackage{hyperref}
\usepackage[usenames,dvipsnames,svgnames,table,rgb]{xcolor}
\hypersetup{				
	unicode=true,           
	pdftitle={Заголовок},   
	pdfauthor={Автор},      
	pdfsubject={Тема},      
	pdfcreator={Создатель}, 
	pdfproducer={Производитель}, 
	pdfkeywords={keywords1} {key2} {key3}, 
	colorlinks=true,       	
	linkcolor=red,          
	citecolor=black,        
	filecolor=black,      
	urlcolor=black           
}

\usepackage{booktabs}
\usepackage{multirow}
\newcommand{\frc}[2]{\raisebox{2pt}{$#1$}\big/\raisebox{-3pt}{$#2$}}
\usepackage{float}

\usepackage{pstool}
 
\usepackage{adjustbox}

\usepackage{tikz}
\usetikzlibrary{shapes.geometric, arrows}

\tikzstyle{cdft} = [rectangle, rounded corners, minimum width=1cm, minimum height=1cm,text centered,text width=2.6cm, draw=black]

\tikzstyle{vfdft} = [rectangle, rounded corners, minimum width=1cm, minimum height=1cm,text centered,text width=2.6cm, draw=red]

\tikzstyle{empty} = [minimum width=1cm, minimum height=1cm,text centered,text width=1cm]

\tikzstyle{arrow} = [thick,->,>=stealth]

\usepackage[justification=centerlast]{caption}
\usepackage{subcaption}
\usepackage[section]{placeins}

\begin{document}
\selectlanguage{english}
\preprint{APS/123-QED}

\title{Adaptive intermolecular interaction parameters for accurate Mixture Density Functional Theory calculations}

\author{Irina Nesterova}%
\email{irina.nesterova@phystech.edu}
\affiliation{%
Moscow Institute of Physics and Technology,\\
Center for Engineering and Technology of MIPT 
}

\author{Yuriy Kanygin}
\email{yuriy.kanygin@phystech.edu}
\affiliation{%
Moscow Institute of Physics and Technology,\\
Center for Engineering and Technology of MIPT 
}

\author{Pavel Lomovitskiy}%
\email{pavel.lomovitskiy@phystech.edu}
\affiliation{%
Moscow Institute of Physics and Technology,\\
Center for Engineering and Technology of MIPT
}

\author{Aleksey Khlyupin}%
\email{khlyupin@phystech.edu}
\affiliation{%
Moscow Institute of Physics and Technology,\\
Center for Engineering and Technology of MIPT
}



\date{\today}

\begin{abstract}

The description of fluid mixtures molecular behavior is significant for various industry fields due to the complex composition of fluid found in nature. Statistical mechanics approaches use intermolecular interaction potential to predict fluids behavior on the molecular scale. The paper provides a comparative analysis of mixing rules applications for obtaining intermolecular interaction parameters of mixture components. These parameters are involved in the density functional theory equation of state for mixtures (Mixture DFT EoS) and characterize thermodynamic mixture properties in the bulk. The paper demonstrates that Mixture DFT EoS with proper intermolecular parameters agrees well with experimental mixtures isotherms in bulk: $Ar + Ne$, $CO_2 + CH_4$, $CO_2 + C_2H_6$ and $CH_4 + C_2H_6$. However, predictions of vapor-liquid equilibrium (VLE) experimental data for $CO_2 + C_4H_{10}$ are not successful. Halgren HHG, Waldman\,--\,Hagler, and adaptive mixing rules that adjust on the experimental data from the literature are used for the first time to obtain intermolecular interaction parameters for the mixture DFT model. The results obtained provide a base for understanding how to validate the DFT fluid mixture model for calculating thermodynamic properties of fluid mixtures on a micro and macro scale.

\end{abstract}
\def\tocname{Content}
\maketitle
\tableofcontents
\newpage
\section{Introduction}

Fluids encountered in nature are rather multicomponent systems, not pure fluids. Therefore, fluid mixture simulations are necessary for designing processes such as separation, enhanced oil recovery (EOR), and others \cite{ potoff2001vapor, heuchel2002molecular, yu2004density, wu2006density, le2015co2, elola2019preferential, cornette2018binary, hofmann1998molecular}. Oil recovery from unconventional reservoirs, where nanopores can constitute about 70\% of the pore volume, is challenging \cite{yu2019compositional}. The behavior of fluid confined in nanopores differs from that of in the bulk. In confinement, forces of solid-fluid and fluid-fluid interactions significantly affect surface phenomena such as capillary condensation, layering transitions, adsorption \cite{ravikovitch1995capillary, bryk2005phase, balbuena1993theoretical}. Besides, even a small concentration of energetically more potent particles can significantly change fluid behavior in confinement \cite{bi2019molecular, wang2018displacement}. To study fluid mixture in pores, it is crucial to consider the following phenomena: selectivity –-- the relation of component concentration in the pore and the bulk and segregation –-- composition difference between fluid near the pore wall and in the pore center \cite{tan1987lennard, tan1992selective, kurniawan2006simulation, roth2009selectivity}. However, we cannot observe such processes at a molecular scale experimentally.

Theory provides an understanding of physical phenomena such as adsorption \cite{balbuena1993theoretical, neimark1998pore, ravikovitch2001characterization, sangwichien2002density}, phase transitions \cite{fu2005vapor,liu2017adsorption, luo2019novel}, capillary condensation \cite{ravikovitch1995capillary, kierlik2002adsorption, neimark2003bridging}, and many others  \cite{telo1980theory, patra1999density, berim2008nanodrop, wu2011classical}, both at molecular and macroscale. Statistical physics approaches build a connection between molecular events with phenomena in confinement and bulk. One of the most commonly used theoretical methods to predict fluid behavior in confinement is Density Functional Theory (DFT) \cite{telo1980theory, balbuena1993theoretical, ravikovitch1995capillary, neimark1998pore, patra1999density,  ravikovitch2001characterization, ravikovitch2001density, kierlik2002adsorption, sangwichien2002density, fu2005vapor, wu2006density, berim2008nanodrop, wu2011classical, aslyamov2017density}. DFT is the rigorous statistical mechanical method, requiring less computational costs than molecular simulation, which can be applied to describe molecular and macroscopic fluid properties \cite{wu2006density}. C. Ebner, W.F. Saam, and D. Stroud were the first to introduce DFT of simple classical fluids in 1976 \cite{ebner1976density}. Later, molecular DFT was developed to account for different molecular interactions by excess Helmholtz free energy terms. To consider short-range repulsion, called a hard spheres interaction, in 1985 Tarazona built Smoothed Density Approximation (SDA) \cite{tarazona1985free}. Afterward, in 1989, another method, i.e. Fundamental Measure Theory (FMT), was founded by Rosenfeld \cite{rosenfeld1989free}. The contribution from long-range attraction is usually treated using Mean Field Approximation (MFA) \cite{tarazona1987phase}. The mixture of Lennard\,--\,Jones (LJ) fluids in terms of DFT, FMT, and MFA was first investigated by Kierlik and Rosenberg in 1991 \cite{kierlik1991density}. Furthermore, the classical DFT was extended for solving particular problems: Statistical Associating Fluid Theory (SAFT) for modeling polymers \cite{chapman1988phase}, Quenched Solid Density Functional Theory (QSDFT) \cite{ravikovitch2006density}, and Random Surface Density Functional Theory (RSDFT) \cite{khlyupin2017random, aslyamov2019theoretical} to take into account the effect of rough surfaces on fluid behavior, and Random Surface Statistical Associating Fluid Theory (RS--SAFT) to study the impact of rough surface on the adsorption of n-alkanes \cite{aslyamov2019random}. However, molecular DFT is not widely used for the prediction of fluid mixture behavior.

Presently, the behavior of fluid mixtures at the molecular scale is mainly studied using Molecular Dynamics (MD) \cite{mecke1999molecular,le2015co2,li2019effects,elola2019preferential}, Statistical Associating Fluid Theory (SAFT) \cite{muller2001molecular,economou2002statistical,liu2017adsorption, herdes2018prediction, liu2019competitive}, Grand Canonical Monte Carlo simulations (GCMC) \cite{cracknell1993grand,palmer2011adsorptive, liu2015adsorption, cornette2018binary}, and Gibbs Ensemble Monte Carlo simulations (GEMC) \cite{pathak2017experimental, bi2019molecular}. However, only a few studies use DFT \cite{sokolowski1990lennard, kierlik1991density, tan1992selective} caused by the number of limitations of the DFT approach. Firstly, only spherical molecules (simple fluids) are modeled by DFT using FMT and MFA, but real molecules have a complex structure that requires other approaches  \cite{liu2017adsorption}. Secondly, it has been shown that DFT calculations deviate from the results of MD and GCMC for mixtures. The discrepancy is explained by applying MFA for attractive interactions of mixture components  \cite{cracknell1993grand, kierlik1991density}.

Therefore, one of the weak sides of DFT consists of attractive interaction descriptions between mixture components. In MFA, effective intermolecular LJ potential determines these interactions depending on the scale and energy parameters $\sigma_{ij}$ and $\epsilon_{ij}$. Thus, for an accurate description of the mixture components interactions, it is necessary to adjust parameters $\sigma_{ij}$ and $\epsilon_{ij}$ of the DFT mixture model. In previous Mixture DFT studies \cite{kierlik1991density, sokolowski1990lennard, tan1992selective, kurniawan2006simulation}, these parameters were found by the Lorentz\,--\,Berthelot (LB) mixing rule. However, the use of this mixing rule is improbable to accurately reproduce the behavior of a real fluid, which is demonstrated in this work, and it was also assumed earlier \cite{kurniawan2006simulation}. It has also been shown for the GCMC simulation in \cite{delhommelle2001inadequacy}. Despite this, the LB mixing rule is widely used in mixture molecular modeling \cite{cracknell1993grand, palmer2011adsorptive, le2015co2, liu2015adsorption, liu2017adsorption, pathak2017experimental, bi2019molecular}. There are also other mixing rules to select the parameters of intermolecular interactions: Halgren HHG (H\,--\,HHG), Waldman\,--\,Hagler (WH), and others \cite{halgren1992representation, waldman1993new, tang1998atoms, schnabel2007unlike}, which allow predicting fluid mixture behavior accurately. Besides, in \cite{al2004generating}, the authors developed adaptive mixing rules, which depends on coefficients, adjusted on the mixture's experimental data in the bulk. There are some works where adaptive mixing rules were applied  \cite{mecke1999molecular, herdes2018prediction, liu2019competitive} in MD, SAFT. It was also used with DFT (LDA + MFA) in \cite{winkelmann2001liquid} to study vapor-liquid interface but LDA poor predicts fluid structure near wall.

(It is also worth noting that all these mixing rules were developed to obtain the intermolecular parameters for the van der Waals equation rather than the Lenard-Jones potential used in the DFT model. Here, the question arises: why can these rules be used for the DFT model? Perhaps, the answer is that both the parameters for the Van der Waals equation and the parameters of the Lenard\,--\,Jones potential describe the same force field. Moreover, that the DFT in the limit on the bulk turns into the equations of state.)

In this paper, we determine intermolecular interaction parameters for mixture components to obtain accurate Mixture DFT EoS. The calculations are performed with classical molecular DFT formulation, where the Rosenfeld FMT version describes hard-sphere interactions and MFA is used for attractive interactions. Different mixing rules are used for the first time to obtain the intermolecular parameters of attractive interaction within the DFT approach. We consider mixing rules  H\,--\,HHG, WH, and adaptive mixing rules: the adaptive Lorenz\,--\,Berthelot mixing rules (ALB) and the adaptive Halgren HHG mixing rules (AH\,--\,HHG). The parameters of the adaptive rules were adjusted to the experimental data of the mixtures in the bulk. Different mixing rules were examined to represent the thermodynamic properties of the mixtures: $Ar + Ne$, $CO_2 + CH_4$, $CO_2 + C_2H_6$, $CH_4 + C_2H_6$, and $CO_2 + C_4H_{10}$. The procedure of $\sigma_{ij}$ and $\epsilon_{ij}$ selection is described in the current work. First, the parameters of the LJ potential for each component of the mixture are determined. Then, the application of LB, H\,--\,HHG, and WH for the mixture thermodynamic properties description was estimated. If these mixing rules were mismanaged, the adaptive mixing rules were accommodated to describe the mixture properties. The Nelder\,--\,Mead optimization method is used to find the parameters $\sigma_{ij}$ and $\epsilon_{ij}$ for the adaptive mixing rules. Finally, with the found intermolecular parameters, we obtain an accurate Mixture DFT EoS. We will use the present work results to describe mixture properties in confinement to study collective adsorption for optimal Enhanced Oil Recovery (EOR) in our future research. We also expect that this work will inspire to extend applications of classical molecular DFT.

The article is organized as follows. First, the DFT model for mixtures is presented. Then, various mixing rules: LB, H\,--\,HHG, WH, ALB, and AH\,--\,HHG are given. We also describe in detail the algorithm of searching intermolecular parameters. The results of applying different mixing rules to describe thermodynamic mixture properties are discussed. Finally, results are summarized in the table reflecting the appropriate mixing rule for describing mixture properties at a particular condition.
\section{Mixture Density Functional Theory}\label{sec:MixDFT}

In this section, we provide the DFT model used in the study. We consider fluid particles in confinement merged with bulk at constant parameters $(T,V,\mu)$. The free energy of the system is Omega potential $\Omega$ that formulates as a functional of the particle distribution function $\rho(\vec{r})$:
\begin{equation}\label{eq:Omega_mix}
\begin{aligned}
    \Omega\left[\rho_1\left(\vec{r}\right), \rho_2\left(\vec{r}\right)\right] &=   F\left[\rho_1\left(\vec{r}\right), \rho_2\left(\vec{r}\right)\right] \\
    & + \sum_{i = 1,2} \int d\vec{r} \rho_i\left(\vec{r}\right)\left(V_i^{ext}\left(\vec{r}\right)-\mu_i\right),
\end{aligned}
\end{equation}
$i$ is the index of component, $F$ is the intrinsic Helmholtz free energy,  $V_i^{ext}$ is the external potential, $\mu_i$ is the chemical potential.

The grand potential $\Omega$ is equal to the minimum value at equilibrium. Thereof we can express the particle distribution function for the component $\rho_i\left(\vec{r}\right)$ as:
\begin{equation}\label{eq:density_mix}
\begin{aligned}
    \rho_i\left(\vec{r}\right) = \rho_i^{bulk}\exp
    \bigg\lbrace -\frac{1}{k_BT}\bigg(
    &\dfrac{\delta F\left[\rho_1\left(\vec{r}\right), \rho_2\left(\vec{r}\right)\right]}{\delta \rho_i\left(\vec{r}\right)} \\
    & + V_i^{ext}\left(\vec{r}\right) - \mu_i^{ex}\bigg)\bigg\rbrace,
\end{aligned}
\end{equation}
where $\rho_i^{bulk}$ the component density in the bulk, $k_B$ the Boltzmann constant. $T$ is the system temperature, $\mu_i^{ex} = \mu_i - \mu_i^{id}$ the excess chemical potential, which will be given below.

In this step, it is essential to formulate Helmholtz free energy functional $F \left[ \rho\right]$, which can be presented as the sum of ideal term $F^{id} \left[ \rho\right]$  and the terms, considering various molecular interactions. For LJ fluid hard-sphere and attractive interactions are to be considered. To treat hard-sphere repulsion $F^{hs} \left[ \rho\right]$, we use FMT \cite{rosenfeld1989free}. $F^{att} \left[ \rho\right]$,which is responsible for the long-range attraction, is considered within the mean field theory framework, as in \cite{ravikovitch2001density}.
\begin{align}
    &F =  F^{id} + F^{hs} + F^{att} \label{eq:sum_mix}\\
    &F^{id} = k_B T \sum_{i = 1,2}\int d\vec{r}\,\rho_i\left(\vec{r}\right)\left(\ln{{(\Lambda_i}^3 \rho_i\left(\vec{r}\right))}-1\right)\label{eq:id_mix}\\
    &F^{hs} = k_B T\int d\vec{r}\,\Phi\left[n_\alpha\left(\rho_1\left(\vec{r}\right), \rho_2\left(\vec{r}\right)\right)\right]\label{eq:HS_mix}\\
    &F^{att}=k_B T \sum_{i,j = 1,2}\iint d\vec{r}\,\rho_i\left(\vec{r}\right)d\vec{r}^\prime\rho_j\left(\vec{r}^\prime\right)U_{ij}^{att}(\vert\vec{r}-\vec{r}^\prime\vert) \label{eq:att_mix}
\end{align}
where $\Lambda_i = \frc{h}{\sqrt{2\pi m_iT}}$ the thermal de Broglie wavelength, $h$ the Planck constant, $m_i$ the mass of the molecule.  $\Phi\left[n_\alpha\left(\rho_1\left(\vec{r}\right), \rho_2\left(\vec{r}\right)\right)\right]$ is the Rosenfeld functional and $n_\alpha$ the weighted density, will be given in Appendix \ref{sec:Appendix_2k_DFT}. The potential of intermolecular interactions $U_{ij}^{att}$ is expressed as:
\begin{equation}
    U_{ij}^{att}\left(r\right)=\ \left\{
    \begin{matrix}
        -\epsilon_{ij}&r<\lambda_{ij}\\
        U_{ij}^{LJ}&\lambda_{ij}<r<r_{cut}\\
        0&r>r_{cut}\\
    \end{matrix}\right.
\end{equation}
\begin{equation}
    U_{ij}^{LJ}=4\epsilon_{ij}\left(\left(\frac{\sigma_{ij}}{r}\right)^{12}- \left(\frac{\sigma_{ij}}{r}\right)^6\right).
\end{equation}
with $r = \vert\vec{r}-\vec{r}^\prime\vert$, $\epsilon_{ij}$ and $\sigma_{ij}$ the effective intermolecular interaction parameters, $\lambda_{ij} = 2^{1/6} \sigma_{ij}$ is the coordinate of LJ minimum, $r_{cut}$ is the cutoff distance, we consider $r_{cut} = \infty$.

\begin{figure}[h]
    \centering
    
    \includegraphics[width=1\linewidth]{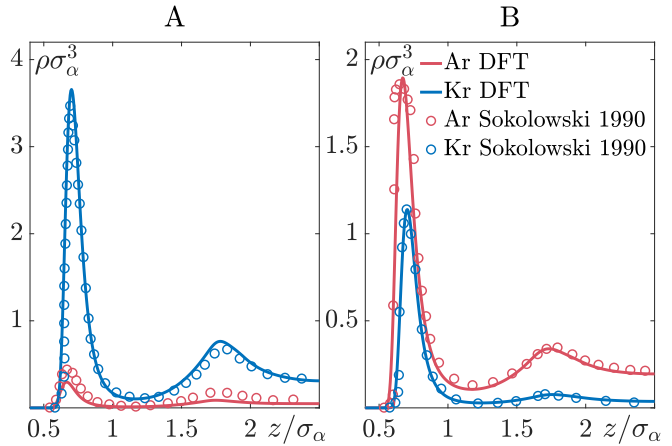}

    \caption{Argon and krypton density profiles in slit-like carbon pore $H = 5\sigma_{\alpha\alpha}$ at $T = 239.6$~K (A) with $\rho_{bulk}^{mix} = 0.444/\sigma_{\alpha\alpha}^3$ and $x_{\alpha} = 0.262$, (B) with $\rho_{bulk}^{mix} = 0.103/{\sigma_{\alpha\alpha}^3}$ and $x_{\alpha} = 0.891$  compared with the results from \cite{sokolowski1990lennard}}
    \label{fig:dft2k_pore}
\end{figure}

The Mixture DFT model results were verified with the simulation \cite{sokolowski1990lennard}, where the Meister--Kroll--Groot version of DFT was used. We calculate equilibrium density profiles for the mixture of argon and krypton in carbon slitlike pore  $H = 5\sigma_{\alpha\alpha}$ at $T = 239.6$~K. LJ potential parameters for argon (index $\alpha$) and krypton (index $\beta$) are given in \cite{sokolowski1990lennard}: $ \sigma_{\alpha\alpha} = 3.405$ \AA, $\frc{\epsilon_{\alpha\alpha}}{k_B} = 119.8$~K, $ \sigma_{\beta\beta} = 3.630$ \AA, $\frc{\epsilon_{\beta\beta}}{k_B} = 163.1$~K. The solid--fluid particle interactions are modeled by 9--3 potential: $U_{sf} = \frc{3^{3/2}\epsilon_{sf}}{2}$
$\left[\left(\frc{\sigma_{sf}}{r}\right)^{9}- \left(\frc{\sigma_{sf}}{r}\right)^3\right]$, with $ \sigma_{\alpha s} = 0.5621\sigma_{\alpha\alpha}$, $\epsilon_{\alpha s} = 9.2367\epsilon_{\alpha\alpha}$, $ \sigma_{\beta s} = 0.5880\sigma_{\beta\beta}$, $\epsilon_{\beta s} = 12.1744\sigma_{\alpha\alpha}$, which were also taken from \cite{sokolowski1990lennard}. We examine the Mixture DFT model on two cases of mixture density and concentration of argon in the bulk: (A) with $\rho_{bulk}^{mix} = 0.444/\sigma_{\alpha\alpha}^3$ and $x_{\alpha} = 0.262$, (B) with $\rho_{bulk}^{mix} = 0.103/{\sigma_{\alpha\alpha}^3}$ and $x_{\alpha} = 0.891$. The Mixture DFT model yields an accurate representation of the components particle distribution functions for different mixtures in the bulk, which are shown in \fref{fig:dft2k_pore}.

Now, we obtain Equation of State for Mixture in the bulk. In the limit of $H \to \infty$, particle distribution function $\rho(\vec{r})$ becomes constant, and the Helmholtz free energy functional turns to the function of variable $\rho$. The expressions for chemical potential and pressure can be derived from Helmholtz free energy equations \ref{eq:sum_mix} -- \ref{eq:att_mix} and given by:

\begin{align}
    &\mu_i = \mu_i^{id} +\mu_i^{hs} + \mu_i^{att}\label{eq:mu_2k}\\
    &\mu_i^{id} = k_B T \ln{\Lambda_i^3 \rho_i}\label{eq:mu_2k_id}\\
    &\mu_i^{hs} = k_B T \left( \frac{\partial \Phi}{\partial n_3} V_i + \frac{\partial \Phi}{\partial n_2} S_i + \frac{\partial \Phi}{\partial n_1} R_i+ \frac{\partial \Phi}{\partial n_0}\right)\label{eq:mu_2k_hs}\\
    &\mu_i^{att} = k_B T \rho_i\int d\vec{r} U_{ii}^{att}(\vec{r}) + k_B T \rho_j\int d\vec{r} U_{ij}^{att}(\vec{r}) \label{eq:mu_2k_att}
\end{align}
with $V_i = \frac{4}{3}\pi R_i^3$, $S_i = \pi R_i^2$, $R_i = \sigma_{ii}/2$ the component particle radius.
\begin{align}
    &p = p^{id} +p^{hs} + p^{att}\label{eq:p_2k}\\
    &p^{id} = \sum_{i = 1,2}\rho_i k_B T \label{eq:p_2k_id}\\
    &p^{hs} = (\rho_1+\rho_2) k_B T \left(\frac{1 + 2\eta+3\eta^2}{(1-\eta)^2}-1\right)\label{eq:p_2k_hs}\\
    &p^{att} = 0.5 \sum_{i,j = 1,2}\rho_i \rho_j k_B T\int d\vec{r} U_{ij}^{att}(\vec{r})\label{eq:p_2k_att}
\end{align}
with $\eta = \sum_{i = 1,2}\rho_i V_i$ is the sum of packing fractions.
Equations \ref{eq:mu_2k_att} and \ref{eq:p_2k_att} rely on $U_{ij}^{att}(\sigma_{ij},\epsilon_{ij})$. Intermolecular interaction parameters $\sigma_{ij}$ and $\epsilon_{ij}$ depend on LJ parameters of pure components. The procedure of searching one component fluid parameters is provided in Appendix \ref{sec:Appendix_1k_search}. When LJ component parameters are known, $\sigma_{ij}$ and $\epsilon_{ij}$ are built as a function of them. It is discussed in detail in the next section.

\section{Mixing Rules}\label{sec:MixRules}

Mixing rules are necessary to determine the parameters $\sigma_{ij}$ and $\epsilon_{ij}$, which characterize the interactions between different types of molecules. Here, $\sigma_{ij}$ is the effective minimum distance between the centers of the molecules, and $\epsilon_{ij}$ is the effective energy well depth of molecular attraction. The Lorentz\,--\,Berthelot rules are usually used to connect $\sigma_{ij}$ and $\epsilon_{ij}$ with pure components parameters. However, this rule does not accurately reproduce mixture properties in the bulk; therefore, other mixing rules have been proposed \cite{halgren1992representation, waldman1993new, al2004generating}. We categorize all rules into two groups: the standard mixing rules (LB, H\,--\,HHG, and WH) and the adaptive mixing rules: (ALB and AH\,--\,HHG), adjusted to the experimental data using the fitting coefficients.

\subsection{Standart Mixing Rules}\label{sec:MixRules_st}

Standard mixing rules are a functional dependence of intermolecular interaction parameters on the parameters of pure components. These rules are easy to use as they do not require additional quantities, such as polarizability, ionization potential, and others, that are difficult to determine \cite{al2004generating}.

\textit{Lorentz\,--\,Berthelot (LB).} The Lorentz\,--\,Berthelot rules are the most popular to obtain intermolecular interaction parameters for molecular simulations. In this rule, the arithmetic mean is used to determine $\sigma_{ij}$, while the geometric mean is used to determine $\epsilon_{ij}$ \cite{al2004generating}.
\begin{equation}\label{eq:Lor-Bert}
    \sigma_{ij} = \frac{\sigma_{ii} + \sigma_{jj}}{2},
    \quad
    \epsilon_{ij} = \sqrt{\epsilon_{ii}\epsilon_{jj}}
\end{equation}

\textit{Halgren HHG (H\,--\,HHG).} 
The Halgren HHG (Harmonic mean of the Harmonic and Geometric mean) rules apply the weighted mean of the arithmetic mean for the definition of $\sigma_{ij}$. The sum of the squares of the component molecule effective diameters is taken as the weighting factor. To calculate $\epsilon_{ij}$, the harmonic mean for the harmonic and geometric mean values of the components are used \cite{halgren1992representation}. The necessity of these rules was motivated by the evidence that the LB mixing rule could not predict the experimental data accurately for rare gases. Thus, new relationships for obtaining intermolecular interaction parameters were proposed to reflect the experimental data better.
\begin{equation}\label{eq:Hal-HHG}
    \sigma_{ij} = \frac{\sigma_{ii}^3 + \sigma_{jj}^3}{\sigma_{ii}^2 + \sigma_{jj}^2},
    \quad
    \epsilon_{ij} = \frac{4\epsilon_{ii}\epsilon_{jj}}{\left(\epsilon_{ii}^{1/2} + \epsilon_{jj}^{1/2}\right)^2}
\end{equation}

\textit{Waldman\,--\,Hagler (WH).}  The Waldman\,--\,Hagler rules use the six power mean  for $\sigma_{ij}$ and the geometric mean of a value $\epsilon\sigma^6$ to determine $\epsilon_{ij}$. It was found that $\epsilon_{ij}$ depends on both $\epsilon$ and $\sigma$ of mixture components \cite{waldman1993new}.
\begin{equation}\label{eq:Wal-Hag}
    \sigma_{ij} = \left( \frac{\sigma_{ii}^6 + \sigma_{jj}^6}{2}\right)^{1/6},
    \quad
    \epsilon_{ij} = \sqrt{\epsilon_{ii}\epsilon_{jj}} \left( \frac{\sigma_{ii}^3 \sigma_{jj}^3 }{\sigma_{ij}^6 }\right)
\end{equation}

\subsection{Adaptive Mixing Rules} \label{sec:MixRules_adap}

Adaptive mixing rules are functionally similar to the standard rules but include fitting coefficients adjusted on the mixture experimental data. They allow analyzing a wide range of values for intermolecular parameters and choosing best to describe a mixture's behavior. The adaptive mixing rules used in this paper are formulated below. Afterward, the algorithm for searching fitting coefficients is given.

\textit{Adaptive Lorentz\,--\,Berthelot (ALB).}
Such formulation for the Lorentz\,--\,Berthelot rule type was proposed by Zudkevitch and Joffe in 1970 \cite{zudkevitch1970correlation}.

\begin{equation}\label{eq:Ad-Lor-Bert}
    \sigma_{ij} = \frac{\sigma_{ii} + \sigma_{jj}}{2} \left(1-k_{ij} \right),
    \quad
    \epsilon_{ij} = \sqrt{\epsilon_{ii}\epsilon_{jj}} \left(1-l_{ij} \right)
\end{equation}
where $k_{ij}$ and $l_{ij}$ are the fitting coefficients for description of $i$ and $j$ component interactions.

 \textit{Adaptive Halgren HHG (AH\,--\,HHG).}
 By analogy with the ALB mixing rule, we propose the adaptive Halgren rule HHG (AH\,--\,HHG), which takes the following form:
 \begin{equation}\label{eq:Ad-Hal-HHG}
    \sigma_{ij} = \frac{\sigma_{ii}^3 + \sigma_{jj}^3}{\sigma_{ii}^2 + \sigma_{jj}^2} \left(1-k_{ij} \right),
    \quad
    \epsilon_{ij} = \frac{4\epsilon_{ii}\epsilon_{jj}}{\left(\epsilon_{ii}^{1/2} + \epsilon_{jj}^{1/2}\right)^2} \left(1-l_{ij} \right)
\end{equation}
here $k_{ij}$ and $l_{ij}$ are the fitting coefficients as in the ALB rule.

\subsection{Algorithm for searching parameters} \label{sec:MixRules_adap_search}

The algorithm's input contains temperature, masses of component molecules and their LJ parameters, the concentration of components in the mixture, and the experimental isotherm or VLE. It is also necessary to set restrictions on the sought coefficients; we use  $k_{ij}, l_{ij} \in (-1,1)$. It was found that the solution for the coefficients is not unique, so one of them can be fixed. We fixed $k_ {ij} = 0$ and varied only one fitting coefficient $l_{ij}$ to get one solution.

The adjustment of the coefficient was carried using the least squares method. As an objective function in the isotherm experimental data case, we use the square of the Mixture DFT EoS pressure deviation from the experimental pressure.
\begin{equation}\label{eq:objective}
    F_{obj} = \frc{1}{n}(\vec{p}-\vec{p}_{exp})\vec{C}(\vec{p}-\vec{p}_{exp})'
\end{equation}
where $\vec{C} = diag \left(\vec{p}_{exp}^{-2}\right)$, $n$ is the length of the experimental data. In equations \ref{eq:p_2k} -- \ref{eq:p_2k_att}, the experimental density values are substituted to calculate $\vec{p}$.

As objective functions while tuning to the VLE data we use: (eq.\ref{eq:objective_vle_1}) the deviation of the vapor pressure according to Mixture DFT EoS from the experimental saturation pressure, (eq.\ref{eq:objective_vle_2}) the deviation of the liquid pressure according to Mixture DFT EoS from the experimental pressure, and (eq.\ref{eq:objective_vle_3}, eq.\ref{eq:objective_vle_4}) the deviation of the chemical potential of the vapor and the liquid component phase calculated according to the Mixture DFT EoS, for each component of the mixture.

\begin{equation}\label{eq:objective_vle_1}
    F^1_{obj} = \frc{1}{n}(\vec{p}^{v}-\vec{p}_{exp})\vec{C_1}(\vec{p}^{v}-\vec{p}_{exp})'
\end{equation}

\begin{equation}\label{eq:objective_vle_2}
    F^2_{obj} = \frc{1}{n}(\vec{p}^{l}-\vec{p}_{exp})\vec{C_2}(\vec{p}^{l}-\vec{p}_{exp})'
\end{equation}

\begin{equation}\label{eq:objective_vle_3}
    F^3_{obj} = \frc{1}{n}(\vec{\mu}^{v}_{1}-\vec{\mu}^{l}_{1})\vec{C_3}(\vec{\mu}^{v}_{1}-\vec{\mu}^{l}_{1})'
\end{equation}

\begin{equation}\label{eq:objective_vle_4}
    F^4_{obj} = \frc{1}{n}(\vec{\mu}^{v}_{2}-\vec{\mu}^{l}_{2})\vec{C_4}(\vec{\mu}^{v}_{2}-\vec{\mu}^{l}_{2})'
\end{equation}
indexes $v,l$ are for the vapor and liquid phase, $\vec{C_1} = \vec{C_2} = diag
\left(\vec{p}_{exp}^{-2} \right)$, $\vec{C_3} = \vec{C_4} = \vec{I}$. In equations \ref{eq:mu_2k} -- \ref{eq:mu_2k_att} and \ref{eq:p_2k} -- \ref{eq:p_2k_att}, the experimental density and the concentration values are substituted to calculate $\vec{\mu}^{phase}_{i}$ and $\vec{p}^{phase}$,

Any optimization approach can carry search for $l_{ij}$. In this work, the Nelder\,--\,Mead method is used. The number of variables is 1. The number of iterations sufficient for the convergence of the method was equal to $10^2$, the reflection parameter $\alpha = 1$, the stretch parameter $\gamma = 2$, the compression parameter $\beta = 0.5$. The $l_{ij}$ is included in equations \ref{eq:mu_2k_att} and \ref{eq:p_2k_att}.
The algorithm's output is the fitted coefficient $l_{ij}$, which determines the effective energy of the intermolecular interactions for mixture components.
\section{Results}\label{sec:results}

Mixture DFT EoS calculations with the use of different mixing rules were performed to describe thermodynamic mixture properties. $Ar + Ne$, $CO_2 + CH_4$, $CO_2 + C_2H_6$, $CH_4 + C_2H_6$, and $CO_2 + C_4H_{10}$ mixtures were considered in separate subsections. In the following pages, Mixture DFT EoS in the case of isotherms will mean equations \ref{eq:p_2k} -- \ref{eq:p_2k_att}, and in the case of VLE equations  \ref{eq:mu_2k} -- \ref{eq:mu_2k_att} and \ref{eq:p_2k} -- \ref{eq:p_2k_att}. For isotherms, we used the objective function  $F_{obj}$ from equation \ref{eq:objective}, and for VLE, the objective functions $F^1_{obj}, F^2_{obj}, F^3_{obj}, F^4_{obj}$ from equations \ref{eq:objective_vle_1} -- \ref{eq:objective_vle_4}.

\begin{figure*}
    \centering
    \includegraphics[width=1\linewidth]{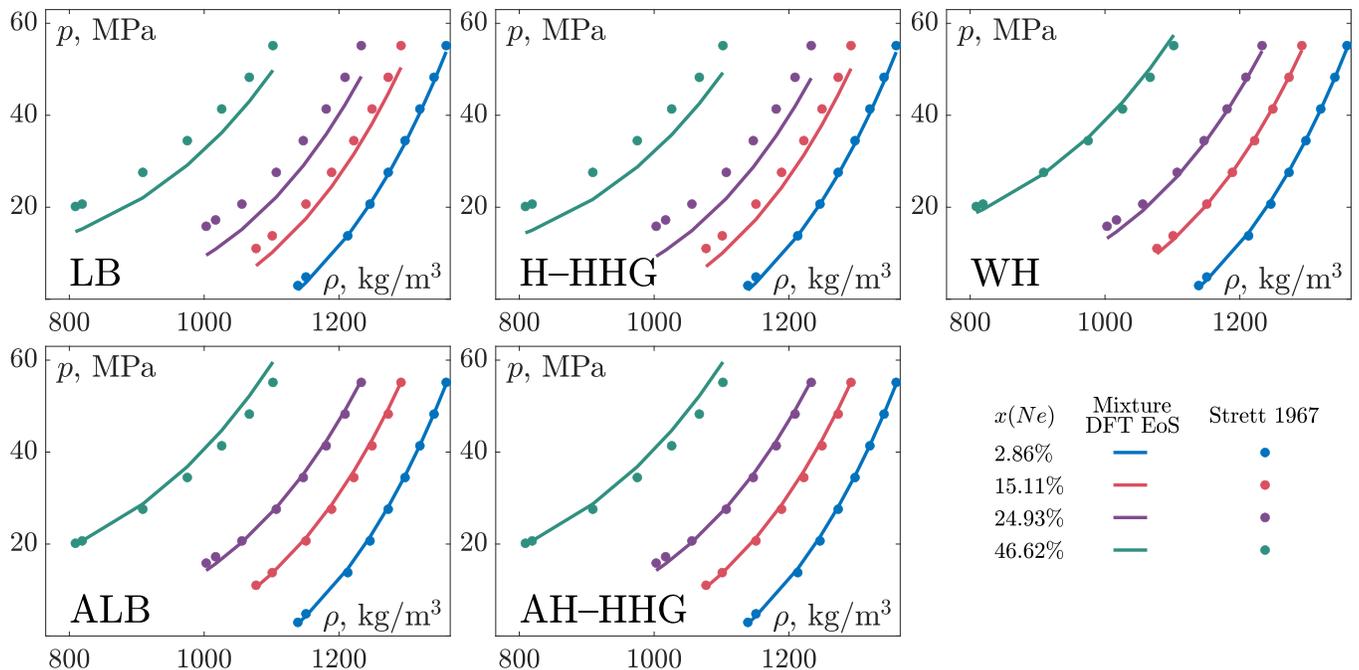}
    \caption{Argon + neon mixture isotherms at $T = 121.36$~K for different mixture composition calculated using DFT and the mixing rule (solid line) in comparison with the experimental data \cite{streett1967liquid}  (circles)}
    \label{fig:an}
\end{figure*}

\begin{figure*}
    \centering
    \includegraphics[width=1\linewidth]{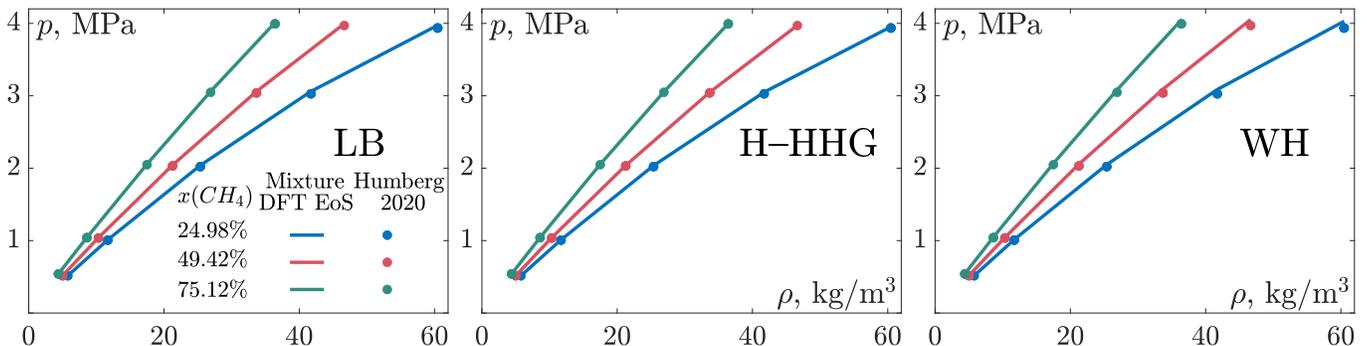}
    \caption{Methane + ethane mixture isotherms at $T = 293$~K for different mixture composition calculated using DFT and the mixing rule (solid line) in comparison with the experimental data \cite{humberg2020measurements}  (circles)}
    \label{fig:me_st}
\end{figure*}

To conclude that the mixing rules together with Mixture DFT EoS manage in describing the isothermal properties of the mixture, we used the criterion for the objective function:
    \begin{equation}\label{eq:criteria}
    F_{obj} < 0.01
    \end{equation}
If the standard mixing rules successfully describe the mixture's thermodynamic properties, the adaptive mixing rules were not applied, as the standard mixing rules are the particular case of adaptive ones with zero coefficients.

\subsection{Argon + Neon}

The mixing rules were verified on isotherms of argon and neon mixture in the liquid phase reproduced in \cite{streett1967liquid}. The mixture is considered at $T = 121.36$~K in a pressure range of up to 60 MPa with different concentrations of the components in the bulk: $2.86\%, 15.11\%, 24.93\%$, and $46.62\%$ neon. First, we found the LJ parameters for the components of the mixture under the considered conditions: for neon $\sigma_{ff} = 2.617$ \AA, $\frc{\epsilon_{ff}}{k_B} = 33.29$~K and for argon $\sigma_{ff} = 3.460$ \AA, $\frc{\epsilon_{ff}}{k_B} = 119.10$~K. Further, various mixing rules were applied to obtain intermolecular parameters, the results are shown in Table \ref{tab:param-an}.

\renewcommand{\arraystretch}{1.1} 
\renewcommand{\tabcolsep}{0.2cm} 
\begin{table}[tb!]
\caption{Intermolecular interaction parameters for $Ar+ Ne$ mixture at $T = 121.36$~K obtained with different mixing rules}
\label{tab:param-an}
\begin{ruledtabular}
\begin{tabular}{ccccc}

\multirow{2}{*}{Mixing Rule} &
  \multirow{2}{*}{$\frc{\epsilon_{ij}}{k_B},$~K} &
  \multirow{2}{*}{$\sigma_{ij},$ \AA} &
  \multirow{2}{*}{$F_{obj}$} &
  \multirow{2}{*}{$l_{ij}$}\\
            &                &                \\
                          \midrule
\multicolumn{1}{l}{$LB$}  & \multicolumn{1}{c}{62.97}          & 3.038        & 0.0355 & -- \\ 
\multicolumn{1}{l}{$H-HHG$}  & \multicolumn{1}{c}{56.98}          & 3.153      &0.0390  & --  \\ 
\multicolumn{1}{l}{$WH$}  & \multicolumn{1}{c}{45.90}          & 3.171       &0.0040  & -- \\ 
\multicolumn{1}{l}{$ALB$}  & \multicolumn{1}{c}{49.21}          & 3.038      & 0.0024   & 0.21851  \\ 
\multicolumn{1}{l}{$AH-HHG$}  & \multicolumn{1}{c}{44.03}          & 3.153       & 0.0024  & 0.22729\\ 
\end{tabular}
\end{ruledtabular}
\end{table}

The isotherms presented in figure \ref{fig:an} show that the mixing rules reproduce the experimental data on the isotherms of the mixture with $2.86\%$ neon successfully; however, with an increase in neon concentration, the equation of state with LB and H–-HHG rules deviate from the experimental data. Mixture DFT EoS using WH, ALB, and AH--HHG rules successfully reproduce systems up to $15.11\%$ neon. For systems with the concentrations of neon $24.93\%$ and $46.62\%$, slight deviations are observed. According to the values of the objective function presented in Table and criterion 27, we concluded that the standard WH rule and the adaptive rules ALB and AH–-HHG could be used to reproduce the isothermal properties of the $Ar + Ne$ mixture accurately.

\begin{figure*}
    \centering
    \includegraphics[width=1\linewidth]{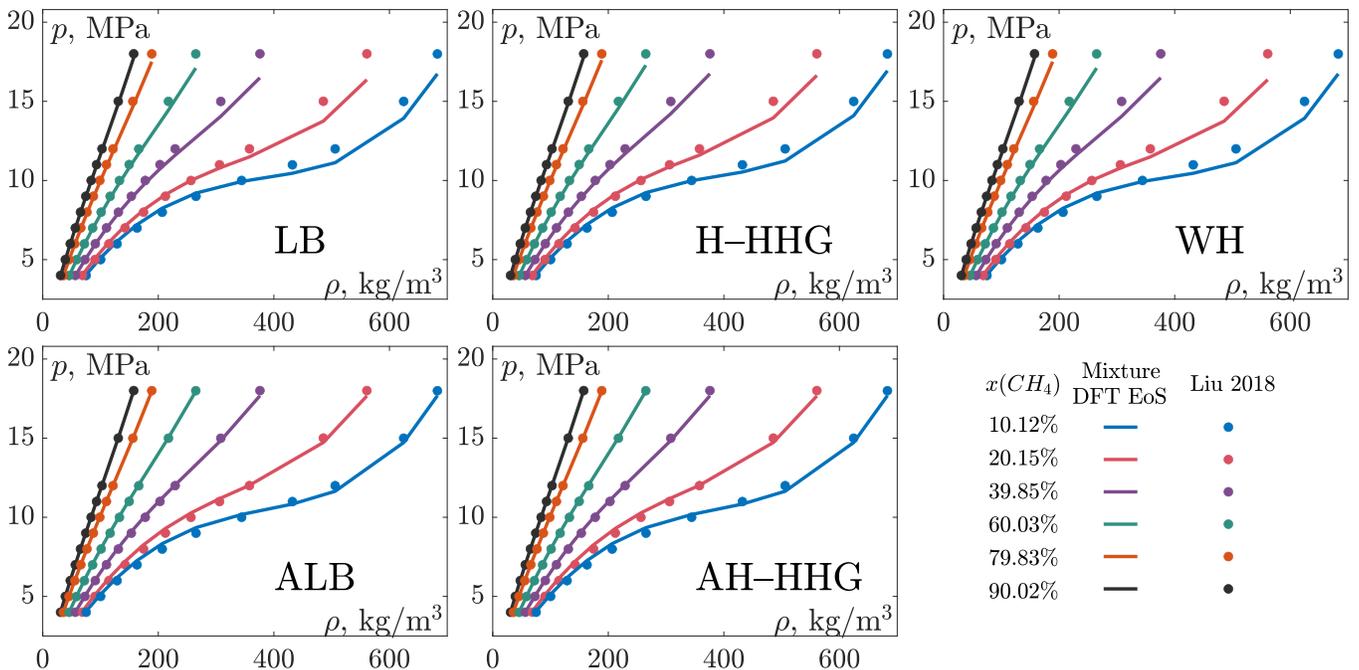}
    \caption{Methane + carbon dioxide mixture isotherms at $T = 313$~K for different mixture composition calculated using DFT and the mixing rule (solid line) in comparison with the experimental data \cite{liu2018density} (circles)}
    \label{fig:mc_isotherm}
\end{figure*}

\subsection{Methane + Ethane}

Methane and ethane gas mixture at $T = 293$~K in a pressure range from 0.5 to 3 MPa with various concentrations of the components: $24.98\%, 49.42\%, 75.12\%$  methane in the mixture in the bulk from [61] was modeled. The parameters of the LJ potential were found for methane $\sigma_{ff} = 3.518$ \AA, $\frc{\epsilon_{ff}}{k_B} = 138.30$~K and ethane $\sigma_{ff} = 4.171$ \AA, $\frc{\epsilon_{ff}}{k_B} = 226.79$~K under considered conditions. The intermolecular parameters obtained with the mixing rules are shown in Table \ref{tab:param-me}.

\renewcommand{\arraystretch}{1.1} 
\renewcommand{\tabcolsep}{0.2cm} 
\begin{table}
\caption{Intermolecular interaction parameters for $CH_4+C_2H_6$ mixture at $T = 293$~K obtained with different mixing rules}
\label{tab:param-me}
\begin{ruledtabular}
\begin{tabular}{cccc}

\multirow{2}{*}{Mixing Rule} &
  \multirow{2}{*}{$\frc{\epsilon_{ij}}{k_B},$~K} &
  \multirow{2}{*}{$\sigma_{ij},$ \AA} &
  \multirow{2}{*}{$F_{obj}$}\\
            &                &                \\
                          \midrule
\multicolumn{1}{l}{$LB$}  & \multicolumn{1}{c}{177.10}          & 3.845       & 4.07e-5\\ 
\multicolumn{1}{l}{$H-HHG$}  & \multicolumn{1}{c}{174.42}          & 3.900        & 1.56e-5  \\ 
\multicolumn{1}{l}{$WH$}  & \multicolumn{1}{c}{156.27}          & 3.911        & 1.83e-4  \\ 
\end{tabular}
\end{ruledtabular}
\end{table}

Figure \ref{fig:me_st} shows that all mixing rules successfully reproduce a mixture of methane and ethane at all considered concentrations, confirmed by the values of the objective function $F_{obj}$ in Table \ref{tab:param-me}, which meets the criteria in eq.\ref{eq:criteria}. The results of the calculation using the H--HHG rule proved to be better than the rest. Based on the results obtained, it can be concluded that all standard mixing rules can be used to describe the behavior of this mixture in the bulk.

\subsection{Methane + Carbon dioxide}

Calculations of the isothermal properties of methane and carbon dioxide mixture in the gas phase \cite{liu2018density} were made to check the mixing rules. A mixture of methane and carbon dioxide is considered at $T = 313$~K  in a pressure range from 3 to 20 MPa with various concentrations of the components:  $10.12\%, 20.15\%, 39.85\%, 60.03\%, 79.83\%, 90.02\%$ methane in a mixture in the bulk. The LJ parameters for the components of the mixture under the considered conditions are as follows: for methane  $\sigma_{ff} = 3.506$ \AA, $\frc{\epsilon_{ff}}{k_B} = 138.16$~K and for carbon dioxide $\sigma_{ff} = 3.512$ \AA, $\frc{\epsilon_{ff}}{k_B} = 219.62$~K. The results are shown in Table \ref{tab:param-mс}.

\begin{figure*}
    \centering
    
    \includegraphics[width=1\linewidth]{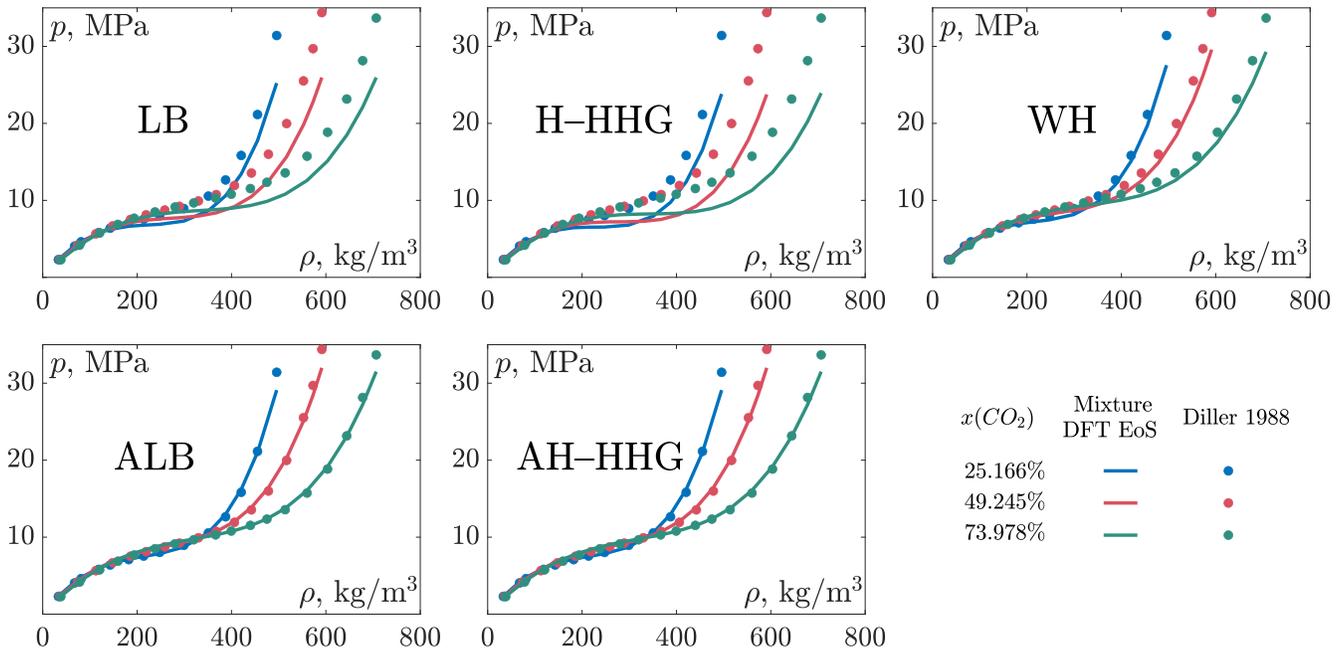}

    \caption{ Ethane + carbon dioxide mixture isotherms at $T = 320$~K for different mixture composition calculated using DFT and the mixing rule (solid line) in comparison with the experimental data \cite{diller1988measurements} (circles)}
    \label{fig:ec_isotherm}
\end{figure*}

\renewcommand{\arraystretch}{1.1} 
\renewcommand{\tabcolsep}{0.2cm} 
\begin{table}[htb!]
\caption{Intermolecular interaction parameters for $CH_4 + CO_2$ mixture at $T = 313$~K obtained with different mixing rules}
\label{tab:param-mс}
\begin{ruledtabular}
\begin{tabular}{ccccc}

\multirow{2}{*}{Mixing Rules} &
  \multirow{2}{*}{$\frc{\epsilon_{ij}}{k_B},$~K} &
  \multirow{2}{*}{$\sigma_{ij},$ \AA} &
  \multirow{2}{*}{$F_{obj}$}&
  \multirow{2}{*}{$l_{ij}$}\\
            &                &                \\
                          \midrule
\multicolumn{1}{l}{$LB$}  & \multicolumn{1}{c}{174.19}          & 3.509        & 9.14e-4 & --\\ 
\multicolumn{1}{l}{$H-HHG$}  & \multicolumn{1}{c}{174.87}          & 3.509        & 6.63e-4 & --\\ 
\multicolumn{1}{l}{$WH$}  & \multicolumn{1}{c}{174.19}          & 3.509         & 9.14e-4 & --\\ 
\multicolumn{1}{l}{$ALB$}  & \multicolumn{1}{c}{163.11}          & 3.509        & 2.43e-4 & 0.06359\\ 
\multicolumn{1}{l}{$AH-HHG$}  & \multicolumn{1}{c}{163.11}          & 3.509        & 2.43e-4  & 0.05097\\ 
\end{tabular}
\end{ruledtabular}
\end{table}

Even though the values of the objective function $F_{obj}$ (see Table \ref{tab:param-mс}) satisfy criterion \ref{eq:criteria} for all mixing rules, visually, see figure \ref{fig:mc_isotherm}, the adaptive rules ALB and AH--HHG coped better with the isothermal properties description. The LB and WH mixing rules gave similar results, and the H--HHG rules worked slightly better than them, but the deviations from the experimental data remained significant. The adaptive rules ALB and AH--HHG show the results better than standard, although small deviations at high concentrations of carbon dioxide in the mixture are preserved. Thus,  to describe this mixture, it will be better to use adaptive mixing rules.

\subsection{Ethane + Carbon dioxide}

To verify the mixing rules, a mixture of ethane and carbon dioxide in the gas phase was simulated \cite{diller1988measurements} at $T = 320$~K in the pressure range from 2 to 35 MPa with various concentrations of the components: $25.166\%, 49.245\%, 73.978\%$ of carbon dioxide in the bulk mixture. First, the parameters of the LJ potential were found for ethane  $\sigma_{ff} = 4.097$~\AA, $\frc{\epsilon_{ff}}{k_B} = 221.15$~K and carbon dioxide $\sigma_{ff} = 3.511$~\AA, $\frc{\epsilon_{ff}}{k_B} = 219.15$~K under the considered conditions. The results for the intermolecular parameters are shown in Table \ref{tab:param-eс}.

\renewcommand{\arraystretch}{1.1} 
\renewcommand{\tabcolsep}{0.2cm} 
\begin{table}[htb!]
\caption{Intermolecular interaction parameters for $C_2H_6 + CO_2$ mixture at $T = 320$~K obtained with different mixing rules}
\label{tab:param-eс}
\begin{ruledtabular}
\begin{tabular}{ccccc}

\multirow{2}{*}{Mixing Rule} &
  \multirow{2}{*}{$\frc{\epsilon_{ij}}{k_B},$~K} &
  \multirow{2}{*}{$\sigma_{ij},$~\AA} &
  \multirow{2}{*}{$F_{obj}$}&
  \multirow{2}{*}{$l_{ij}$}\\
            &                &                \\
                          \midrule
\multicolumn{1}{l}{$LB$}  & \multicolumn{1}{c}{220.15}          & 3.804         & 0.0229 & -- \\ 
\multicolumn{1}{l}{$H-HHG$}  & \multicolumn{1}{c}{220.15}          & 3.849         &0.0421 & -- \\ 
\multicolumn{1}{l}{$WH$}  & \multicolumn{1}{c}{198.49}          & 3.859         &0.0043 & -- \\ 
\multicolumn{1}{l}{$ALB$}  & \multicolumn{1}{c}{198.55}          & 3.804         & 7.35e-4 &  0.03600\\ 
\multicolumn{1}{l}{$AH-HHG$}  & \multicolumn{1}{c}{191.69}          & 3.849         & 7.35e-4 & 0.12918\\ 
\end{tabular}
\end{ruledtabular}
\end{table}

Figure 5 demonstrates that the standard WH mixing rule describes intermolecular interactions much better than the LB and H--HHG rules. However, deviations from the experimental data at high concentrations of $CO_2$ remain and grow with an increase in pressure. The values of the objective function $F_{obj}$ for WH, ALB, and AH--HHG satisfy criteria \ref{eq:criteria}, see Table \ref{tab:param-eс}, but the adaptive rules coped better. We can conclude that to describe the ethane and carbon dioxide mixture isothermal properties, the adaptive mixing rules ALB and AH--HHG should be utilized.

\subsection{Butane + Carbon dioxide}

In this section, we perform calculations of the phase equilibria of a mixture using Mixture DFT EoS. The phase diagram of carbon dioxide and butane mixture at $T = 311.09$~K, obtained experimentally in \cite{niesen1989vapor+}, was considered. The LJ parameters for carbon dioxide are $\sigma_{ff} = 3.517$~\AA, $\frc{\epsilon_{ff}}{k_B} = 219.91$~K, for butane are $\sigma_{ff} = 5.268$~\AA, $\frc{\epsilon_{ff}}{k_B} = 369.50$~K. Table \ref{tab:param-cb} shows the mixture's intermolecular parameters under the considered conditions.

\begin{figure} [!h]
    \begin{subfigure}{\linewidth}
    \includegraphics[width=1\linewidth]{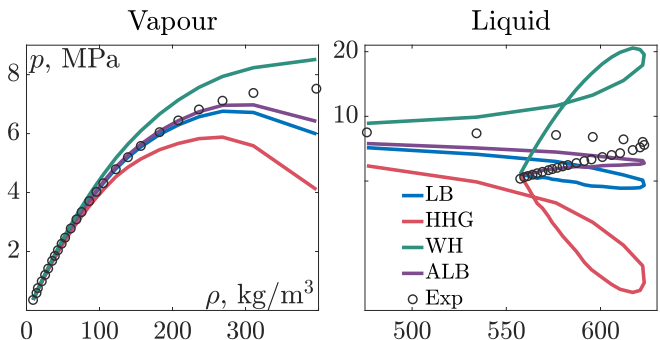}
    \end{subfigure}
    \caption{Butane + carbon dioxide mixture VLE at $T = 311.09$~K calculated using DFT and different mixing rules (solid line) in comparison with the experimental data \cite{niesen1989vapor+}  (circles)}
    \label{fig:cb}
\end{figure}

\begin{figure*} [!t]
    \begin{subfigure}{\linewidth}
    \includegraphics[width=0.95\linewidth]{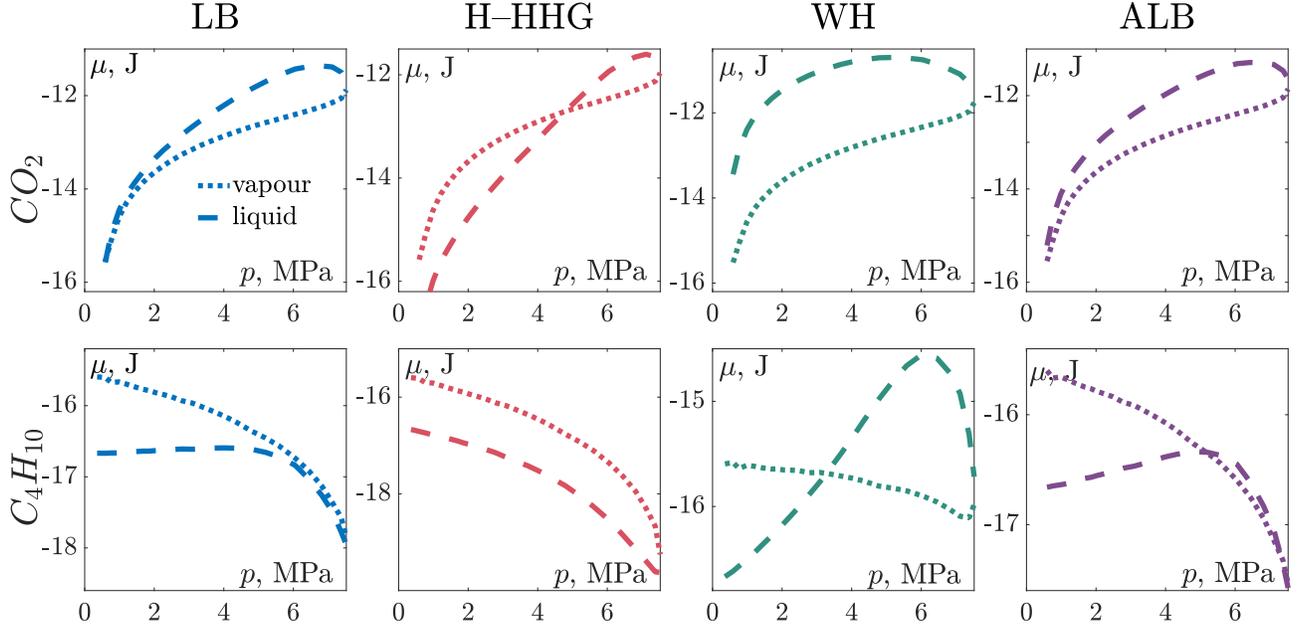}
    \end{subfigure}
     \caption{Chemical potentials of carbon dioxide and butane in vapor (points) and liquid (dashes line) states calculated using DFT and different mixing rules for VLE at $T = 311.09$~K}
    \label{fig:cb_mu}
\end{figure*}

Figures \ref{fig:cb} and \ref{fig:cb_mu} show the results of calculating the mixture pressure and the chemical potentials of the mixture components using Mixture DFT EoS and the standard mixing rules: LB, H--HHG, and WH. Table \ref{tab:cb_st_obj} shows the objective functions \ref{eq:objective_vle_1} -- \ref{eq:objective_vle_4} results for Mixture DFT EoS with the considered mixing rules. The results obtained using standard mixing rules agree with the experimental data for vapor up to 3 MPa. At higher densities, the deviations are observed, particularly: the WH rule overestimate vapour pressure, while LB and H--HHG rules underestimate it. As for liquid pressure, the H-HHG and WH rules show non-physical results (loops are formed) and it is confirmed with the high values of $F_{obj}^2$ in Table \ref{tab:cb_st_obj}. The results on the chemical potentials of two phases: liquid and vapor, the LB rule worked best here and only for carbon dioxide, judging by the values of the objective function $F_{obj}^3$ from Table \ref{tab:cb_st_obj}. These results demonstrate that standard mixing rules work well only for certain conditions.

\renewcommand{\arraystretch}{1.1} 
\renewcommand{\tabcolsep}{0.2cm} 
\begin{table}[htb!]
\caption{Intermolecular interaction parameters for $C_4H_{10} + CO_2$ mixture at $T = 311.09$~K obtained with different mixing rules}
\label{tab:param-cb}
\begin{ruledtabular}
\begin{tabular}{ccc}

\multirow{2}{*}{Mixing Rule} &
  \multirow{2}{*}{$\frc{\epsilon_{ij}}{k_B},$~K} &
  \multirow{2}{*}{$\sigma_{ij},$ \AA} \\
            &                &                \\
                          \midrule
\multicolumn{1}{l}{$LB$}  & \multicolumn{1}{c}{285.06}          & 4.393          \\ 
\multicolumn{1}{l}{$H-HHG$}  & \multicolumn{1}{c}{280.31}          & 4.728    \\ 
\multicolumn{1}{l}{$WH$}  & \multicolumn{1}{c}{155.85}          & 4.760      \\ 
\multicolumn{1}{l}{$ALB$}  & \multicolumn{1}{c}{270.35}          & 4.393      \\ 
\end{tabular}
\end{ruledtabular}
\end{table}

\renewcommand{\arraystretch}{1.1} 
\renewcommand{\tabcolsep}{0.2cm} 
\begin{table}[htb!]
\caption{Objective functions $F^1_{obj}$--$F^4_{obj}$ values obtained by standard mixing rules: LB, H--HHG, WH}
\label{tab:cb_st_obj}
\begin{ruledtabular}
\begin{tabular}{ccccc}

\multirow{2}{*}{Mixing Rule} &
  \multirow{2}{*}{$F_{obj}^1$} & \multirow{2}{*}{$F_{obj}^2$} &
  \multirow{2}{*}{$F_{obj}^3$} & \multirow{2}{*}{$F_{obj}^4$}\\
            &                &                \\
                          \midrule
\multicolumn{1}{l}{$LB$}  & \multicolumn{1}{c}{0.0024}          & 0.8364 & 0.3370 & 0.4113         \\ 
\multicolumn{1}{l}{$HHG$}  & \multicolumn{1}{c}{0.0153}          & 15.1753 & 0.7654 &  1.1245   \\
\multicolumn{1}{l}{$WH$}  & \multicolumn{1}{c}{0.0051}          & 10.5947 & 3.5276 & 0.6575      \\ 
\end{tabular}
\end{ruledtabular}
\end{table}

\renewcommand{\arraystretch}{1.1} 
\renewcommand{\tabcolsep}{0.2cm} 
\begin{table*}
\caption{Objective functions $F^1_{obj}$--$F^4_{obj}$ values obtained by optimization}
\label{tab:cb_ad_obj}
\begin{ruledtabular}
\begin{tabular}{cccccc}

\multirow{2}{*}{Optimized  $F_{obj}$} &
  \multirow{2}{*}{$F_{obj}^1$} & \multirow{2}{*}{$F_{obj}^2$} &
  \multirow{2}{*}{$F_{obj}^3$} & \multirow{2}{*}{$F_{obj}^4$} & \multirow{2}{*}{$l_{ij}$} \\
            &                &                \\
                          \midrule
\multicolumn{1}{l}{$F_{obj}^1$}  & \multicolumn{1}{c}{0.0006}          & 0.6034 & 1.0018 &  0.3291 & 0.1157      \\ 
\multicolumn{1}{l}{$F_{obj}^2$}  & \multicolumn{1}{c}{0.0009}          & 0.1207 & 0.6177 & 0.3356 & 0.0629    \\ 
\multicolumn{1}{l}{$F_{obj}^3$}  & \multicolumn{1}{c}{0.0049}          & 3.2971 & 0.2055 &  0.5354   & -0.0727    \\ 
\multicolumn{1}{l}{$F_{obj}^4$}  & \multicolumn{1}{c}{0.0006}          & 0.3321 & 0.8579 &  0.3268 & 0.0979      \\ 
\multicolumn{1}{l}{$\sum_i F_{obj}^i$}  & \multicolumn{1}{c}{0.0012}          & 0.1533 & 0.5384 &  0.3438 & 0.0491      \\ 
\multicolumn{1}{l}{$\sum_i \tilde F_{obj}^i$}  & \multicolumn{1}{c}{0.0012}          & 0.1428 & 0.5519 &  0.3422 & 0.0516      \\ 
\end{tabular}
\end{ruledtabular}
\end{table*}

Since standard mixing rules cannot describe the VLE of the mixture well, we  use the adaptive ALB rule. In the case of VLE, the goal is to achieve both mechanical equilibrium, i.e., the equality of pressures of a mixture of vapor and liquid, and chemical equilibrium, i.e., the equality of chemical potentials of vapor and liquid for each component of the mixture. To this end, we consider the objective functions $F_{obj}^1$--$F_{obj}^4$ from equations \ref{eq:objective_vle_1}--\ref{eq:objective_vle_4}, where $F_{obj}^1$ and $F_{obj}^2$ reflect the mechanical equilibrium, $F_{obj}^3$ and $F_{obj}^4$ reflect the chemical equilibrium.

While searching for the optimal coefficient $l_{ij}$ with different objective functions, the values of the $l_{ij}$ obtained vary. Table \ref{tab:cb_ad_obj} shows the values of the objective functions and the optimal coefficient $l_{ij}$ depending on the optimization of a specific objective function pointed in the first column. It was found that the objective functions are not compatible with each other; when one of them is optimized, the values of the others worsen. The table shows that the optimization of $F_{obj}^3$ increases the rest of the objective functions dramatically. Pairwise, $F_{obj}^1$ and $F_{obj}^4$, $F_{obj}^2$ and $F_{obj}^4$  appeared to be in good agreement, i.e., their optimal coefficients are close. When optimizing the sum of objective functions, the result turned to be similar to the optimization result of $F_{obj}^2$. It might happen due to the highest values of this objective function at the deviation from the experimental data.

We also tried to optimize the sum of objective functions normalized to their worst value during the optimization of each of them, i.e., $\sum_i \tilde F_{obj}^i = \sum_i F_{obj}^i/a_i$, where $a_1 = 0.0049, a_2 = 3.2971, a_3 = 1.0018, a_4 = 0.5354$, however, in this case, fair values were obtained only for $F_{obj}^1$ objective function. The attempt to optimize two coefficients $k_{ij}$ and $l_{ij}$ gives a set of solutions with the same values of the objective functions. As a result, the simultaneous optimization of the objective functions describing the mechanical and chemical equilibrium is challenging. However, Mixture DFT EoS with ALB rules coped better than standard mixing rules with VLE curve representation (see Figure \ref{fig:cb}). ALB rules concede only LB rules in $CO_2$ chemical potential description. Though, the calculated VLE curve still differs significantly from the experimental one. 

As a result, neither standard nor adaptive mixing rule could help Mixture DFT EOS represent the VLE of the considered mixture. The ALB rule best describes the pressure-- density relation, and the LB rule best characterizes the chemical equilibrium of carbon dioxide.

While studying the previous DFT works, we noticed that the VLE calculations usually differ from the experimental data: the critical temperature calculated using DFT usually overestimates the experimental value, and the density of the liquid phase predicted by DFT underestimates the experimental one \cite{kierlik1991density, ravikovitch2001density, winkelmann2001liquid}. In the paper \cite{ravikovitch2001density} the authors mentioned the difficulty of simultaneously achieving chemical equilibrium and equal pressure of the liquid and vapor phases for a pure substance using DFT. In the articles \cite{kierlik1991density, winkelmann2001liquid}, the deviations of the calculated VLE were explained by the mean-field approach, which poorly describes the influence of the attraction of interaction of fluid particles on their properties in the bulk. Thus, calculating vapor-liquid equilibrium remains a difficult task for DFT.
\section{Conclusion}\label{sec:out&concl}

\begin{table*}
\caption{Results for mixing rules application. Mixture DFT EOS with the mixing rule "$+$" successfully, "$+/-$" satisfactory reproduce fluid properties; "$-$" failed to reproduce fluid properties, "$\times$" - were not considered.}
\label{tab:mixing_rules}

\begin{tabular}{|c|c|c|c|c|c|c|c|c|c|}
\hline
Mixture & Data & Phase & $T,K$ & Value & LB & H--HHG & WH & ALB & A--HHG\\
\hline
$Ar+Ne$ & Isotherm & Liquid & 121.36 & $p$ & $-$ & $-$ & $+$ & $+$ & $+$ \\
\hline
$CH_4+C_2H_6$ & Isotherm & Vapour & 293 & $p$ & $+$ & $+$ & $+$ & $+$ & $+$ \\
\hline
$CO_2+CH_4$ & Isotherm & Vapour & 313 & $p$ & $+/-$ & $+/-$ & $+/-$ & $+$ &
$+$ \\
\hline
$CO_2+C_2H_6$ & Isotherm & Vapour & 320 & $p$ & $-$ & $-$ & $+/-$ & $+$ & $+$ \\
\hline
$CO_2+C_4H_{10}$ & VLE & Vapour & 311.09 & $p$ & $+/-$ & $-$ & $+/-$ & $+/-$ & $ \times$ \\
\hline
$CO_2+C_4H_{10}$ & VLE & Liquid & 311.09 & $p$ & $-$ & $-$ & $-$ & $+/-$ & $ \times$ \\
\hline
$CO_2+C_4H_{10}$ & VLE & V+L & 311.09 & $\mu_{CO_2}$ & $+/-$ & $+/-$ & $-$ & $+/-$ & $ \times$ \\
\hline
$CO_2+C_4H_{10}$ & VLE & V+L & 311.09 & $\mu_{C_4H_{10}}$ & $+/-$ & $-$ & $+/-$ & $+/-$ & $ \times$ \\
\hline
\end{tabular}
\end{table*}

Previously, the Mixture DFT approach proved to be insufficiently accurate compared to MD and GCMC for describing the behavior of a fluid mixture in a pore. The inaccuracy was explained by using the mean field approach to describe the attractive interactions of fluid molecules \cite{cracknell1993grand, kierlik1991density}. Earlier, to obtain the parameters of intermolecular interaction, the LB mixing rule was used, which, as was shown in this work, does not accurately reproduce the thermodynamic properties of mixtures in the bulk. In this work, a comparison was made of various mixing rules for describing a binary mixture's behavior. Besides, we present an approach to adjust the parameters of intermolecular interaction for a mixture DFT model according to the experimental data. The algorithm was tested on a set of mixtures: $Ar + Ne$, $CO_2 + CH_4$, $CO_2 + C_2H_6$, $CH_4 + C_2H_6$, $CO_2 + C_4H_{10}$. 

As a result, it was revealed that among all the mixtures considered, the LB rule makes a good description of the isothermal properties only for the mixture of hydrocarbons, where any other standard or adaptive mixing rule is also suitable. The WH rule can be used to describe a mixture of rare gases, while for other mixtures, it is better to apply the adaptive ALB or AH--HHG rules. Table \ref{tab:mixing_rules} shows the results of how mixing rules cope with describing the properties of various mixtures. We did not obtain VLE results that agree well with the experimental data. We hope that this work will provide an impetus for developing and applying the classical DFT approach for modeling mixtures in the confinement. In the future, we plan to apply the Mixture DFT approach to study the competitive adsorption of oil components for the analysis and prediction of EOR in the oil and gas industry.
\appendix

\section{Details of Mixture DFT}\label{sec:Appendix_2k_DFT}

Here, we give details of Mixture DFT.

$F^{hs}\left[\rho\right]$ in eq. \ref{eq:HS_mix} contains Rosenfeld functional $\Phi\left[n_\alpha\left(\rho\left(\vec{r}\right)\right)\right]$,   \cite{rosenfeld1989free}, which is given:
\begin{eqnarray}\label{eq:rosienfield}
    \Phi =  -n_0 \ln{\left(1-n_3\right)} & + & \frac{n_1 n_2- \vec{n_1}\cdot\vec{n_2}}{1-n_3} \nonumber\\  & & +  \frac{n_2^3-3n_2 \vec{n_2}\cdot\vec{n_2}}{24\pi\left(1-n_3\right)^2},    
\end{eqnarray}

Functions $n_\alpha, \bm{n}_\beta$ are weighted densities ($\alpha = 0,1,2,3;\,\beta = 1,2$):
\begin{equation}\label{eq:weighted_dens}
    n_\alpha \left(\vec{r}\right)= \sum_i \int d^3r^\prime \rho_i\left(\vec{r}^\prime\right)\omega_\alpha^i\left(\vec{r}-\vec{r}^\prime\right),
\end{equation}
where $\omega_\alpha^i, \bm{\omega}_\beta^i$ are the weight functions of $i$ component; $\omega_3^i\left(\vec{r}\right)=\theta\left(R_i-r\right)$, $\omega_2^i\left(\vec{r}\right)=\delta\left(R_i-r\right)$, ${\vec{\omega}}_2^i\left(\vec{r}\right)=\frac{\vec{r}}{r}\delta\left(R_i-r\right)$, $\omega_1^i= \frac{\omega_2}{4\pi R_i}$, $\omega_0^i = \frac{\omega_2}{4\pi R_i^2}$, ${\vec{\omega}}_1^i= \frac{{\vec{\omega}}_2}{4\pi R_i}$, $\delta$ and $\theta$ are the Dirac delta function and the Heaviside step function, respectively, $R_i$ is $i$ component particle radius.\\

\section{One component DFT}\label{sec:Appendix_1k_DFT}

DFT calculations were also performed for one component fluid model and compared with the results from \cite{ravikovitch2001density}. Here, we consider nitrogen at $T = 77.4$~K and relative pressure $\frc{p}{p_0} = 0.7$, where $p_0 = 101860$~Pa is the saturation pressure at this temperature. The parameters of fluid-fluid interactions $\sigma = 3.758$ \AA, $\frc{\epsilon_{ff}}{k_B} = 105.29$~K were obtained by the algorithm described.

\begin{figure}[!b]
    \centering
    
    \includegraphics[width=1\linewidth]{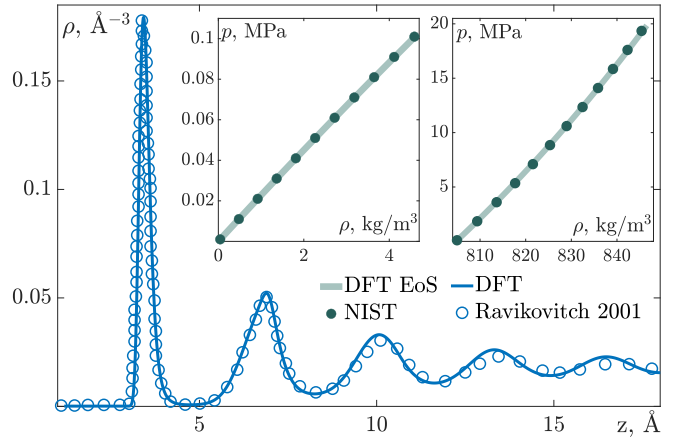}

    \caption{Nitrogen density profile in slit-like carbon pore $H = 10\sigma_{ff}$ at $T = 77.4$~K and relative pressure $p/p_0 = 0.7$ compared with the results from \cite{ravikovitch2001density}. Inset: nitrogen bulk isotherm at $T = 77.4$~K calculated using DFT in comparison with the data from NIST Chemistry WebBook (on the left for the vapor phase, on the right for the liquid phase)}
    \label{fig:dft1k_pore}
\end{figure}

The chemical potential and the pressure of one component fluid can be found as follows:

\begin{align}
    &\mu = \mu^{id} +\mu^{hs} + \mu^{att}\label{eq:mu_1k}\\
    &\mu^{id} = k_B T \ln{\Lambda^3 \rho}\label{eq:mu_1k_id}\\
    &\mu^{hs} = k_B T \left( \frac{\partial \Phi}{\partial n_3}  V + \frac{\partial \Phi}{\partial n_2} S + \frac{\partial \Phi}{\partial n_1} R + \frac{\partial \Phi}{\partial n_0}\right)\label{eq:mu_1k_hs}\\
    &\mu^{att} = k_B T \rho\int d\vec{r} U^{att}(\vec{r}) \label{eq:mu_1k_att}
\end{align}
with $V = \frac{4}{3}\pi R^3$, $S = \pi R^2$, $R = \sigma_{ff}/2$~--- particle radius.

\begin{align}
    &p = p^{id} + p^{hs} + p^{att}\label{eq:p_1k}\\
    &p^{id} = \rho k_B T\label{eq:p_1k_id}\\
    &p^{hs} = \rho k_B T \left(\frac{1 + 2\eta+3\eta^2}{(1-\eta)^2}-1\right)\label{eq:p_1k_hs}\\
    &p^{att} = 0.5 \rho^2 k_B T\int d\vec{r} U^{att}(\vec{r}) \label{eq:p_1k_att}
\end{align}

Equations \ref{eq:mu_1k_hs}, \ref{eq:mu_1k_att} and \ref{eq:p_1k_hs}, \ref{eq:p_1k_att} depend on $\sigma_{ff}$ and $\epsilon_{ff}$. These parameters characterize a particular fluid at particular conditions. Below, in \ref{sec:Appendix_1k_search}, we present the procedure to find fluid LJ parameters.\\

\section{Fluid parameters searching procedure} \label{sec:Appendix_1k_search}

In this section, we describe the procedure for selecting the parameters of pure components. The selection of the parameters was done according to the experimental data on the fluid's isothermal properties in the bulk. The experimental data were taken from the NIST Chemistry WebBook.

The algorithm's input consists of temperature, molecular mass, an array of experimental data on pressure-density dependence. Also, boundary conditions should be set at the input of the algorithm to search for parameters. To determine the boundary values, if they are unknown, one can draw a field of values of the objective function, which will be given below, and visually determine the parameters' search area.

The square of the DFT pressure deviation (\ref{eq:p_1k} -- \ref{eq:p_1k_att}) from the experimental value is used as an objective function. An array of experimental data on density is substituted into equations (\ref{eq:p_1k} -- \ref{eq:p_1k_att}). The optimization process is performed by the parameters $\sigma_{ff}$ and $\epsilon_{ff}$, directly involved in equations \ref{eq:p_1k_hs} and \ref{eq:p_1k_att}. To find the parameters $\sigma_{ff}$ and $\epsilon_{ff}$, the Nelder\,--\,Mead method is used with the following parameters: the number of variables is 2; the number of iterations sufficient for the convergence of the algorithm is $10^4$, the reflection parameter $\alpha = 1$, the stretching parameter $\gamma = 2$, and the compression parameter $\beta = 0.5$. We obtain LJ parameters $\sigma_{ff}$ and $\epsilon_{ff}$ of pure fluid at the algorithm's output. For example, for nitrogen from the example above, we received the parameters:$\sigma = 3.758$ \AA, $\frc{\epsilon_{ff}}{k_B} = 105.29$~K, at a temperature $T = 77.4$~K in the system.
\bibliography{references}

\end{document}